\documentclass{article}

\usepackage{microtype}
\usepackage{graphicx}
\usepackage{booktabs} 

\usepackage{hyperref}

\usepackage{subfig}

\usepackage[accepted]{icml2020}

\usepackage{multirow}

\usepackage[colorinlistoftodos]{todonotes}

\usepackage{xspace}

\newcommand{\gat}{\textsc{GAT}\xspace}
\newcommand{\gmn}{\textsc{GMN}\xspace}

\newcommand{\llgnn}{\textsc{LL-GNN}\xspace}

\newcommand{\coatt}{\textsc{MHCADDI}\xspace}
\newcommand{\decagon}{\textsc{Decagon}\xspace}
\newcommand{\decagonoh}{\textsc{Decagon-OH}\xspace}
\newcommand{\decagonfp}{\textsc{Decagon-FP}\xspace}
\newcommand{\decagonr}{\textsc{Decagon-R}\xspace}
\newcommand{\fppred}{\textsc{FP-Pred}\xspace}

\newcommand{\drugbank}{\textsc{DrugBank} \xspace}
\newcommand{\drugcombo}{\textsc{DrugCombo} \xspace}

\newcommand{\mlgnn}{\textsc{Bi-GNN}\xspace}

\usepackage{amsmath}
\usepackage{amssymb}
\usepackage{amsthm}
\usepackage{bm}
\usepackage{algorithm}
\usepackage{algorithmic}

\usepackage{bbm}

\icmltitlerunning{Bi-Level Graph Neural Networks for Drug-Drug Interaction Prediction}

\begin{document}

\twocolumn[
\icmltitle{Bi-Level Graph Neural Networks for Drug-Drug Interaction Prediction}



\icmlsetsymbol{equal}{*}

\begin{icmlauthorlist}
\icmlauthor{Yunsheng Bai}{equal,ucla}
\icmlauthor{Ken Gu}{equal,ucla}
\icmlauthor{Yizhou Sun}{ucla}
\icmlauthor{Wei Wang}{ucla}
\end{icmlauthorlist}

\icmlaffiliation{ucla}{Department of Computer Science, University of California, Los Angeles, USA}

\icmlcorrespondingauthor{Yunsheng Bai}{yba@ucla.edu}
\icmlcorrespondingauthor{Ken Gu}{kengu13@ucla.edu}

\icmlkeywords{Machine Learning, ICML}

\vskip 0.3in
]



\printAffiliationsAndNotice{\icmlEqualContribution} 


\section{Bi-Level Graph Neural Networks}
\label{sec-intro}


We introduce \mlgnn \emph{\textbf{}} (\emph{\underline{Bi}}-Level \emph{\underline{G}}raph \emph{\underline{N}}eural \emph{\underline{N}}etworks) for modeling biological link prediction tasks such as drug-drug interaction (DDI)~\cite{Vilar2012} and protein-protein interaction (PPI)~\cite{keskin2008prism}. Taking drug-drug interaction as an example, existing methods using machine learning either only utilize the link structure between drugs without using the graph representation of each drug molecule~\cite{Zitnik2018,gae}, or only leverage the individual drug compound structures without using graph structure for the higher-level DDI graph~\cite{deac2019drugdrug}. Readers are referred to a recent survey~\cite{ddi_survey} for more details.

We demonstrate our model using the drug-drug interaction prediction task as an example. Drug-drug interactions occur when the presence of one drug changes the effect of another drug producing an observed side effect~\cite{drugbank}. Specifically, we consider the transductive setting of drug repurposing~\cite{pushpakom2019drug}, in which predicting interactions between existing drugs along with leveraging interaction data can help infer similar physiological effects for other existing drugs~\cite{zhou2015drug}. This is especially important with the recent COVID-19 pandemic, as drug repurposing techniques aim to find existing drugs that may be helpful in treading COVID-19~\cite{hamiltoncovidtalk}.  
Our framework is also extendable to other biological link prediction tasks with different interacting biological entities, e.g. proteins~\cite{protein_graph1, protein_graph2}.

\begin{figure}
\centering
\includegraphics[width=0.48\textwidth]{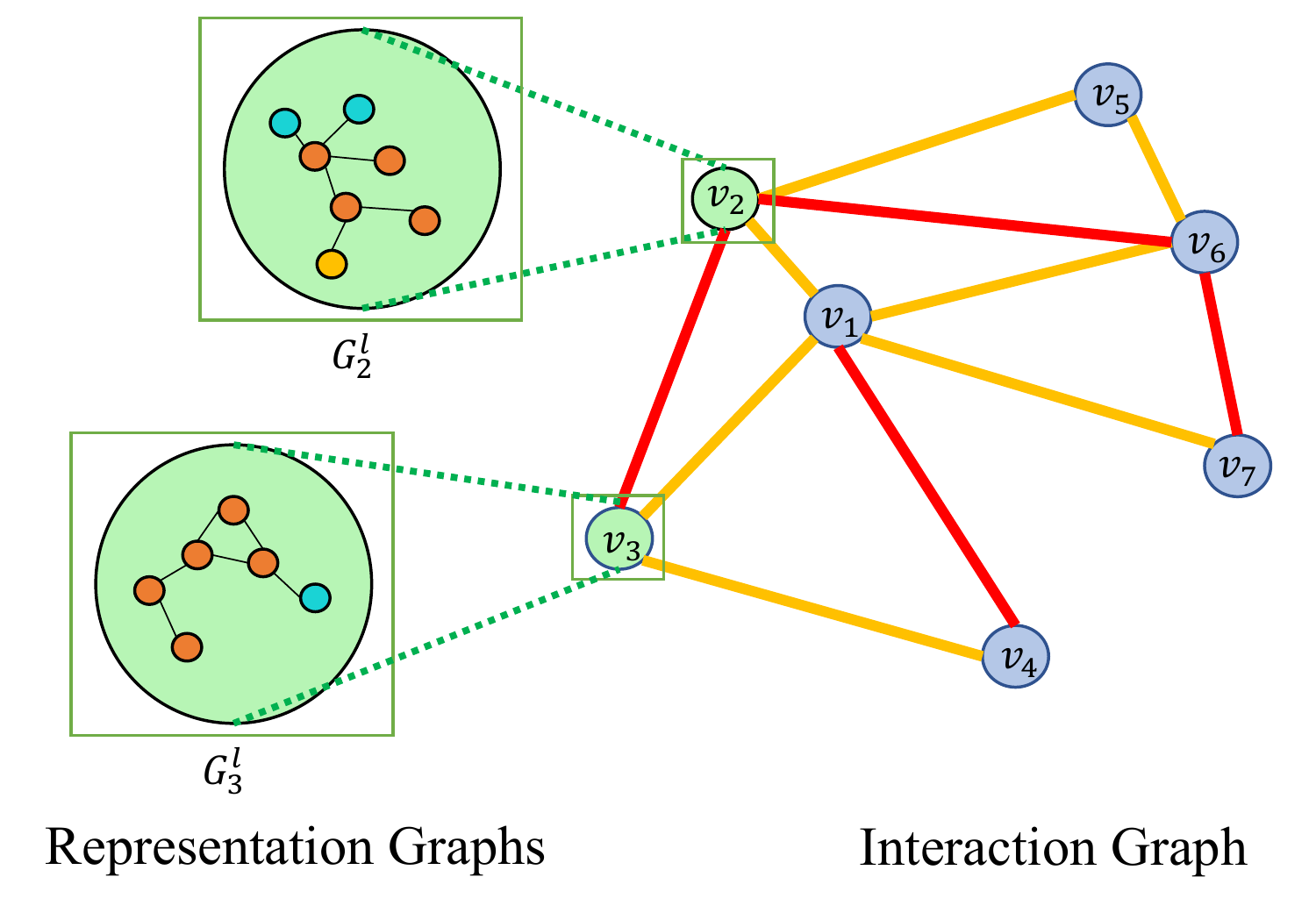}
\vspace*{-6mm}
\caption{A bi-level graph view of the drug-drug interaction data, in which the number of levels is 2. 
Existing GNN methods operate only on either the representation graphs or the single interaction graph without utilizing both under the GNN framework.
Node colors in the representation graphs denote molecular level element types.
Edge colors in the interaction graph denote drug interactions types. }
\label{fig:multilevelgraph}
\vspace*{-6mm}
\end{figure}

\begin{table*}[h]
    \centering
    \caption{Comparison of baseline methods. $\gat^l$ runs on the representation graphs while $\gat^h$ runs on the interaction graph}
    \begin{tabular}{l|lll}
        \multirow{1}{*}{\textbf{Method}}
         &  \textbf{Low-Level Model} & \textbf{High-Level Model} & \textbf{Feature of Drugs}\\
         \hline
         \fppred 
         & N/A & N/A & Molecular Structure \\     
         
         \llgnn 
         & $\textsc{GAT}^l \times 5 $ & N/A & Molecular Structure \\     
         
         \coatt~\cite{deac2019drugdrug}
         & $\textsc{\{GAT-GMN\}}^l \times 3$ & N/A & Molecular Structure \\

         \decagonr~\cite{Zitnik2018} 
         & N/A & $\textsc{GAT}^h \times 3 $ & Random \\

         \decagonfp~\cite{Zitnik2018} 
         & N/A & $\textsc{GAT}^h \times 3 $ & Molecular Structure \\

        \decagonoh~\cite{Zitnik2018} 
         & N/A & $\textsc{GAT}^h \times 3 $ & Learned Transductively \\

        \mlgnn (this paper)
         & $\textsc{GAT}^l \times 5 $ & $\textsc{GAT}^h \times 3 $ & Molecular Structure \\
    \end{tabular}
    \label{table:methods}
\vspace*{-6mm}
\end{table*}

The key idea is to fundamentally view the data as a bi-level graph, in which the highest level is a representing the interaction between biological entities (interaction graph), and each biological entity itself is further expanded to its intrinsic graph representation (representation graphs), in which the graph is either flat like a drug compound \cite{molecular_graph} or hierarchical like a protein with amino acid level graph, secondary structure, tertiary structure, etc~\cite{wiki:Protein}. 
Our \mlgnn not only allows the usage of information from both the high-level interaction graph and the low-level representation graphs, but also offers a baseline for future research opportunities to address the bi-level nature biological interaction networks. 

\textbf{Definitions} \enspace We denote an undirected, unweighted graph $\mathcal{G}=(\mathcal{V},\mathcal{E})$ with $N=|\mathcal{V}|$ nodes. In addition, we let ${\mathcal{G}^h}=(\mathcal{V}^h, \mathcal{E}^h)$ denote the higher level interaction network with node set $\mathcal{V}^h$ and edge set $\mathcal{E}^h$. Moreover each $e^h_i \in \mathcal{E}^h$ can be of different types. Let $\mathcal{R}$ denote the set of all edge types. Each ${v_i} \in \mathcal{V}^h$ is also a graph which we denote  as ${\mathcal{G}^l_i}=(\mathcal{V}^l_i, \mathcal{E}^l_i)$. The set of all graphs in the lower level is $\{\mathcal{G}^l_1, \mathcal{G}^l_2, ..., \mathcal{G}^l_{|\mathcal{V}^h|}\}$. Fig~\ref{fig:multilevelgraph} shows an example with $|\mathcal{V}^h|= 7$.

Node features for the $i^{th}$ representation graph are summarized in a $N_i\times D_l$ matrix $\bm{H^{l}_i}$ where $N_i=|\mathcal{V}_i|$. Likewise node features for the interaction graph are denoted by a $N\times D_h$ matrix $\bm{H^{h}}$. Taking DDI as an example in which the representation graphs are molecular structures of drugs, $H^l_i$ would represent atom features such as the atomic number, whether an atom is aromatic, its hybridization, the number of hydrogen atoms attached to the atom etc. of $\mathcal{G}^l_i$. Currently, we do not consider edge features in the lower level graph but this can be easily extended in future work.

\textbf{\mlgnn} \enspace \mlgnn consists of the following sequential stages: 1) \textit{Lower Level Representation Graph Neural Network} generates vector representations for each representation graph; 2) \textit{Higher Level Interaction Graph Neural Network} further propagates information from the lower level graph embeddings to neighboring nodes in the interaction graph, which provides the final graph representations to a \textit{fully connected network} to obtain a final link prediction score. In this scenario, the representation graph may contain more than one level, one at the amino acid level and another at the secondary structure level.

\subsection{Stage I:  Lower Level Representation Graph Embedding}\label{ssec:1.1}


In Stage I, we generate the representation graphs' graph embeddings. In general this can be obtained through any graph embeddinging method. For example, one can use node embedding models such as Graph Convolutional Networks \cite{kipf2017semi}, Graph Attention Networks \cite{velickovic2018graph}, or Graph Isomorphism Networks \cite{xu2018how} followed by a $\mathrm{READOUT}$ function, a function that takes the embeddings of nodes as input and outputs a single embedding for the graph.  Additionally, one can use hierarchical graph representation models such as $\textsc{DiffPool}$ \cite{ying2018hierarchical} or $\textsc{Graph-U-Net}$ \cite{graphunet}. 

In this work, we use Graph Attention Networks (GAT) with multi-scale $\mathrm{READOUT}$ of the updated node embeddings. Formally, the $k$-th layer of GAT is defined as: 
\begin{equation} \label{eq:gat} \bm{x_i}^{(k+1)} = \mathrm{ReLU}(\alpha_{i,i}\hat{\bm{x_i}}^{(k)} + \sum_{j \in \mathcal{N}(i)} \alpha_{i,j}\hat{\bm{x_j}}^{(k)}) \end{equation}
\begin{equation} \label{eq:gat_attention} \alpha_{i,j} = \dfrac{\mathrm{exp}(\mathrm{LeakyReLU}({\bm{a}^{(k)}}^T [\hat{\bm{x_i}}^{(k)} || \hat{\bm{x_j}}^{(k)}]))}{\sum_{m \in \mathcal{N}(i)} \mathrm{exp}(\mathrm{LeakyReLU}({\bm{a}^{(k)}}^T [\hat{\bm{x_i}}^{(k)} || \hat{\bm{x_m}}^{(k)}])) }. \end{equation}
Here, $\hat{\bm{x_i}}$ is the transformed node embedding from initial feature:
\begin{equation} \label{eq:init_transformation} \hat{\bm{x_i}} = \bm{W}\bm{x_i}.
\end{equation}
$\mathcal{N}(i)$ is the set of all first-order neighbors of node $i$ plus node $i$ itself; $\bm{W}^{(l)} \in \mathbb{R}^{D^{l} \times D^{l+1}}$ is the weight matrix associated with the $n$-th GAT layer;  $\alpha_{i,j}$ is a scalar attention weight that node i gives to node j; $\bm{a}^{(l)} \in \mathbb{R}^{2D^{l}}$ is the attention weight vector; and $ || $ is vector concatenation.

\textbf{Multi-Scale READOUT} \enspace Inspired by GIN~\cite{xu2018how}, we concatenate the node representations across all layers of the GAT. This allows the model to consider all structural information, at various levels of granularity. For $K$ GAT layers the representation graph embeddings $\bm{x_G}$ can be expressed as:
\begin{equation} \bm{x_G} = \mathrm{CONCAT}(\mathrm{READOUT}(\bm{x^{(k)}_v} | v \in G) | k = 1,2, ... K). \end{equation} 

\subsection{Stage II: Higher Level Interaction Node Embedding}\label{ssec:1.2}

Graph embeddings from Stage I as initial node features $\bm{H^{h}} \in \mathbb{R}^{N \times D_h}$ for the interaction graph. This is motivated by the intuition that the lower level network presents a useful initial representation from which the higher level network further enhances for the task. 
Similar to Stage I, many different node embedding methods may be used to refine the entity representation. We continue to use a different set of \gat layers for the higher level node embedding. Because only the final node representation in the interaction graph is needed, there is no multi-scale $\mathrm{READOUT}$. 
Additionally, in the case of multiple edge types, the \gat propagation becomes 
\begin{equation} \label{eq:gat} \bm{x_i}^{(t+1)} = \mathrm{ReLU}(\sum_{r\in \mathcal{R}}(\alpha^r_{i,i}{{\hat{\bm{x_i}}}^{r(t)}} + \sum_{j \in \mathcal{N}^r(i)} \alpha^r_{i,j}{{\hat{\bm{x_j}}}^{r(t)}})) \end{equation}
where $\mathcal{N}^r(i)$ is the set of neighbors of node $i$ that have edge type $r$; $\hat{\bm{x}}^r_i = \bm{W}_r\bm{x^r_i}$; and $\alpha^r_{i,j}$ is calculated as in Equation~\ref{eq:gat_attention} but parameterized by $\bm{a}_r$.

For link prediction, we concatenate the representations of a pair of entities after $T$ upper level \gat layers and feed it into fully connected layers to predict a link prediction score: 
\begin{equation}\mathrm{pred}(v_i,v_j) = \mathrm{MLP}(\bm{x_i} || \bm{x_j}).
\end{equation}
We choose this decoder over simpler alternatives such as dot product between drug embeddings, because empirically we find this yields better performance.

For multiclass link prediction (multiple edge types), the fully connected layers output $C$ scores for $C$ classes. We use cross entropy loss with logits for the loss function.

\section{Experiments}
\label{sec-exp}


We train \mlgnn on two DDI datasets, \drugbank~\cite{biosnapnets,drugbank} and \drugcombo~\cite{drugcombo} whose details can be found in the supplementary material.

\subsection{Baseline Methods}
To fairly compare the uses of different levels of information we aim to use similar architectures as \mlgnn, summarized in Table~\ref{table:methods} and detailed in the supplementary material.

\textbf{Representation Graph Models} \enspace
For models that only use the lower level representation graph we focus on three variants. The first which we call \fppred, feeds extended connectivity fingerprint (ECFP)~\cite{ECFP} representations directly into the prediction layers. ECFPs are representations of chemical structures which capture relevant molecular features and molecular structure. The second is \llgnn which is the lower level model outlined in section \ref{ssec:1.1}. The last model is \coatt  \cite{deac2019drugdrug}
which uses intra-graph message passing and inter-graph co-attention. We implement a similar model that uses Graph Matching Networks (GMN) \cite{li2019graph} for inter-graph attention. 



\textbf{Interaction Graph Models} \enspace 

Under models that only use the interaction graph, we consider \decagon~\cite{Zitnik2018}, which considers protein-protein networks and protein-drug networks on top of the DDI as the input graph. We adopt \decagon for the DDI network only. More specifically, we focus on three different initializations of $\bm{H^{h}} \in \mathbb{R}^{N \times D_h}$, the input drug features: (1) one hot initialization which we call \decagonoh, (2) random initialization denoted as \decagonr, and  (3) ECFP~\cite{ECFP} initialization which we refer to as \decagonfp. 

\begin{table}[h]
\caption{Overall link prediction accuracy of all the methods on two datasets. Top two results are highlighted in bold.}

    \begin{tabular}{l|ll ll}
      \multirow{2}{*}{\textbf{Method}} & 
      \multicolumn{2}{c}{\textbf{\drugbank}} &
      \multicolumn{2}{c}{\textbf{\drugcombo}} \\ 
      & {ROC} & {PR} 
      & {ROC} & {F1} \\
    \hline
    \fppred
    & 0.630 & 0.647
    & 0.648 & 0.495 
    \\
      \hline
      \llgnn 
      & 0.822 & 0.800 
      & 0.813 & 0.661
      \\
      \coatt
      & 0.864 & 0.844 
      & 0.824 & 0.681
      \\
      \hline

      \decagonr 
      & 0.898 & 0.896 
      & 0.853 & \textbf{0.753}
      \\
      \decagonfp 
      & 0.905 & 0.903 
      & 0.844 & 0.731
      \\
      \decagonoh 
      & \textbf{0.940} & \textbf{0.939}
      & \textbf{0.864} & 0.748
      \\
     \hline
      \mlgnn 
      & \textbf{0.933} &  \textbf{0.929}
      & \textbf{0.859} & \textbf{0.767}
      \\
    \end{tabular}
\label{table-acc}
\end{table}


\begin{figure}[h]
    \centering
    \includegraphics[width=0.5\textwidth]{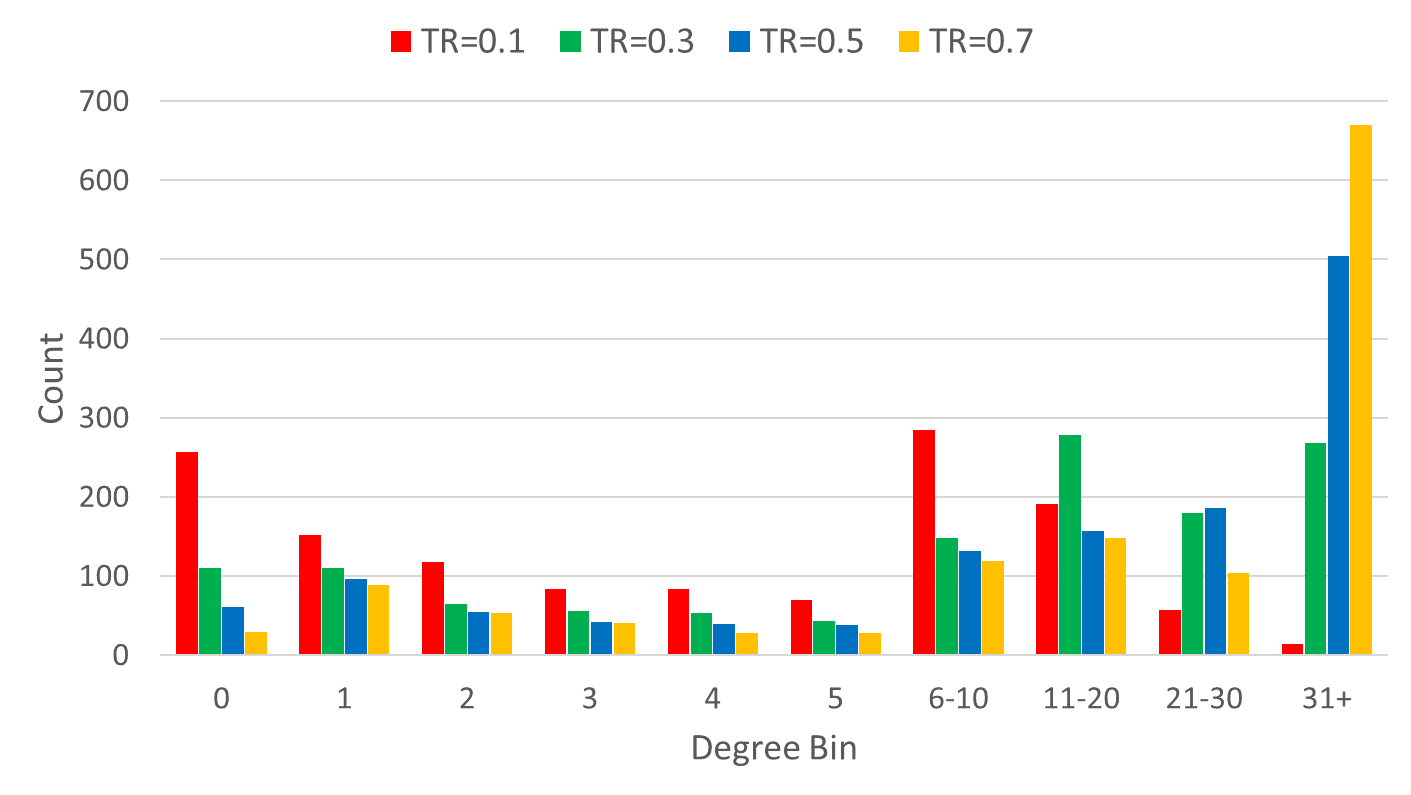}
    \caption{The distribution of node degrees across different training data ratios (percentage of DDI pairs in the training set denoted as ``TR''). 
    }
    \label{fig:degree_bin_counts}
\vspace*{-6mm}
\end{figure}

\begin{figure*}
\centering
\subfloat[TR=0.1]{\includegraphics[scale=0.155]{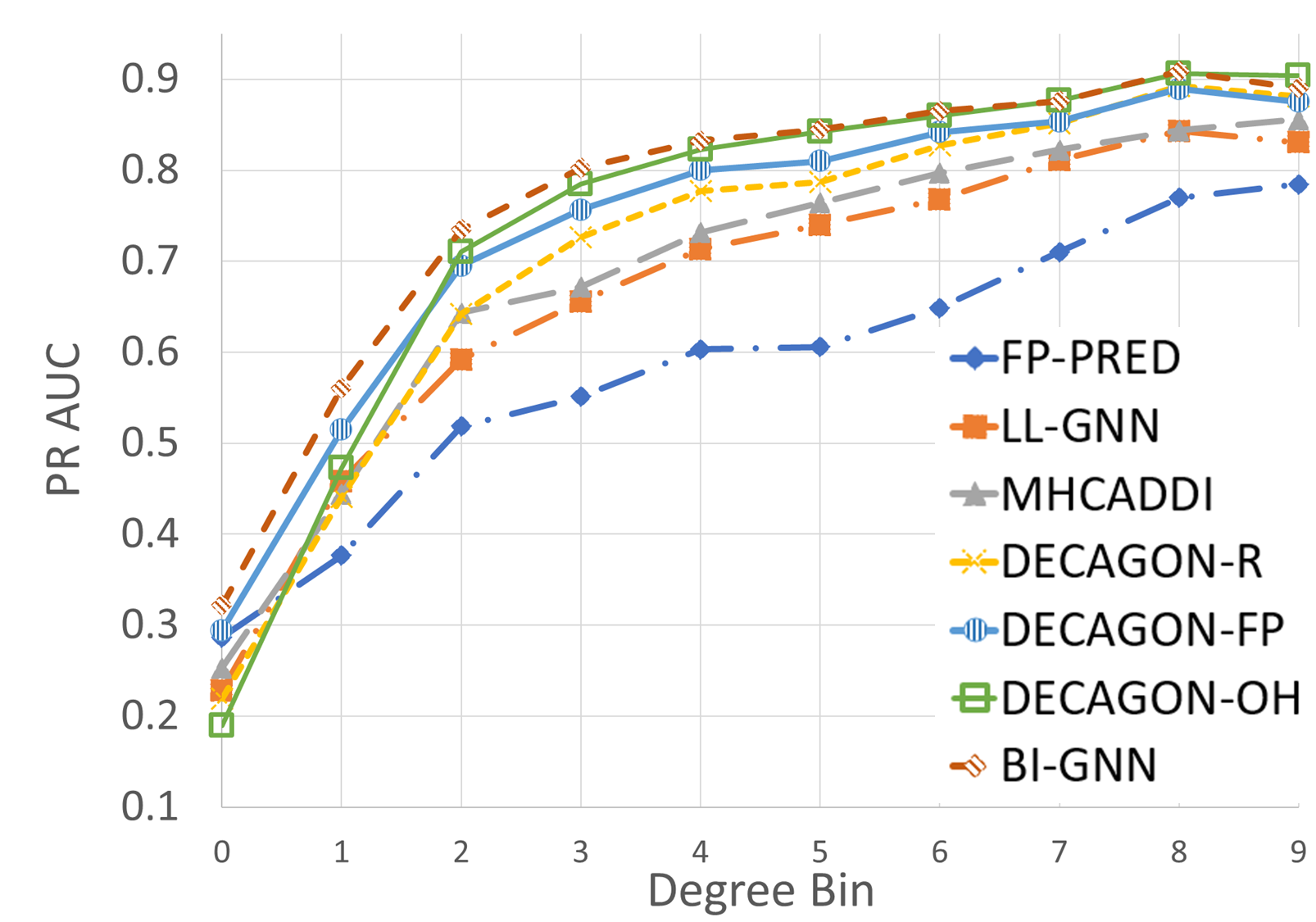}}
\subfloat[TR=0.3]{\includegraphics[scale=0.13]{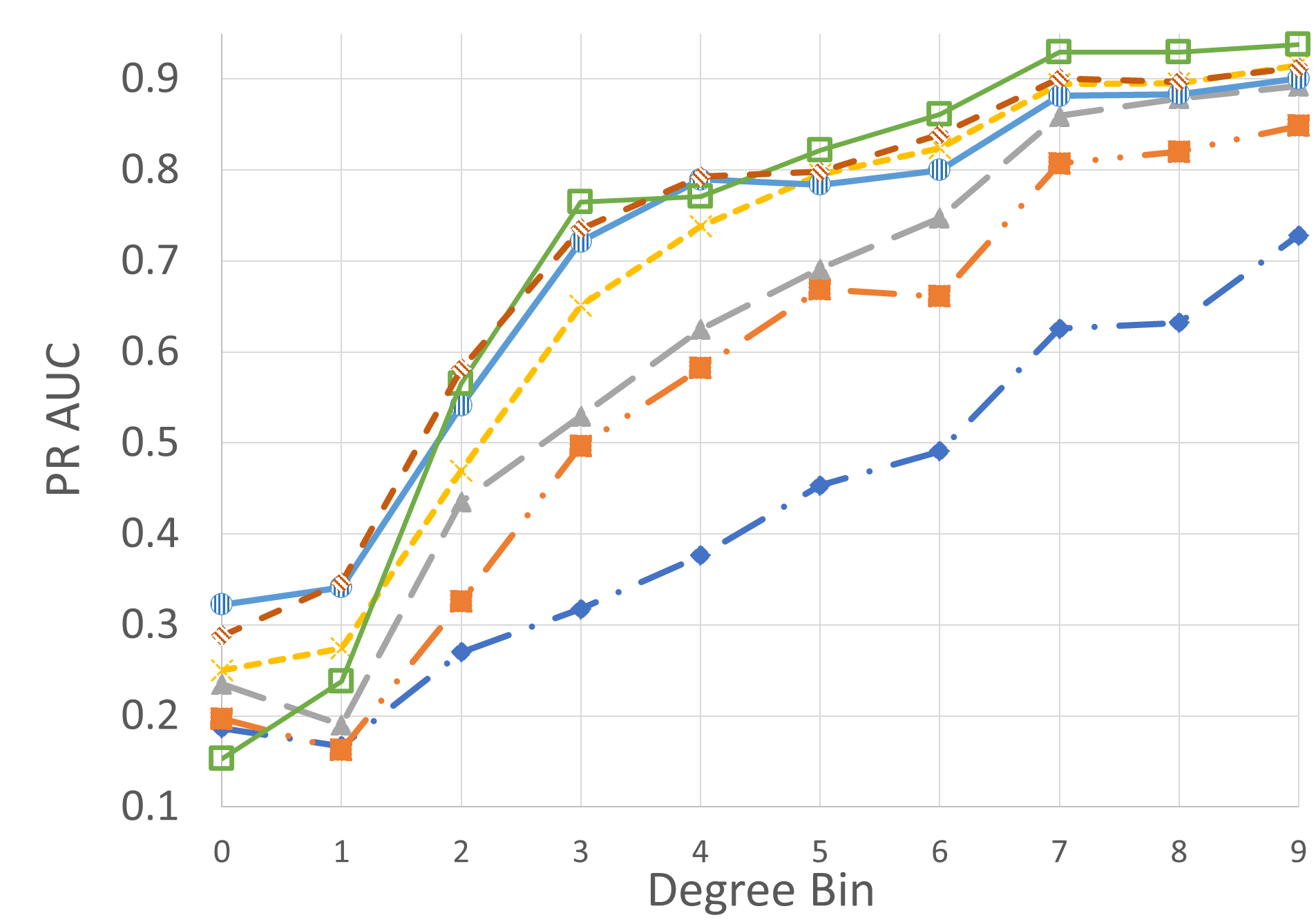}}
\subfloat[TR=0.5]{\includegraphics[scale=0.13]{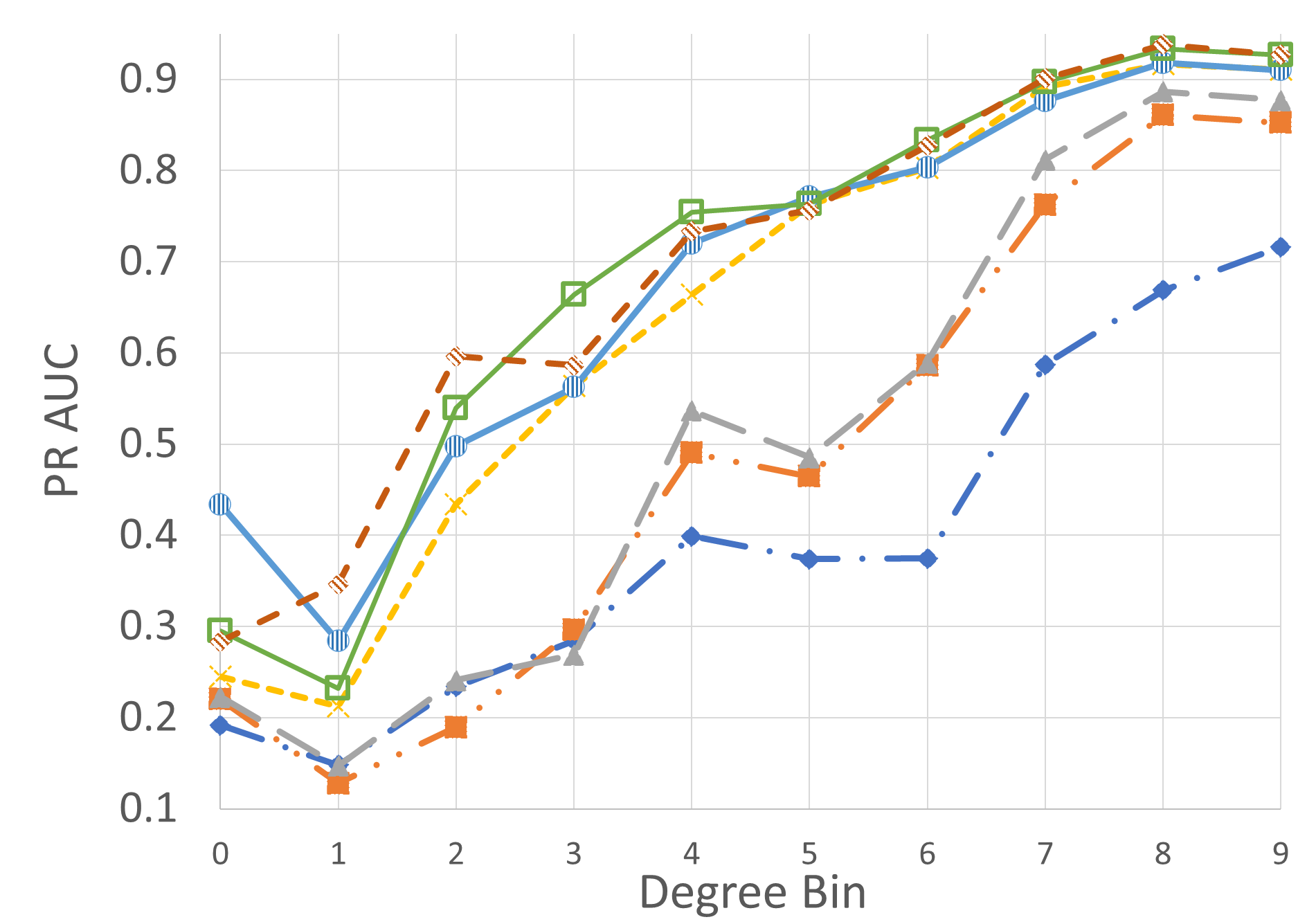}}
\subfloat[TR=0.7]{\includegraphics[scale=0.13]{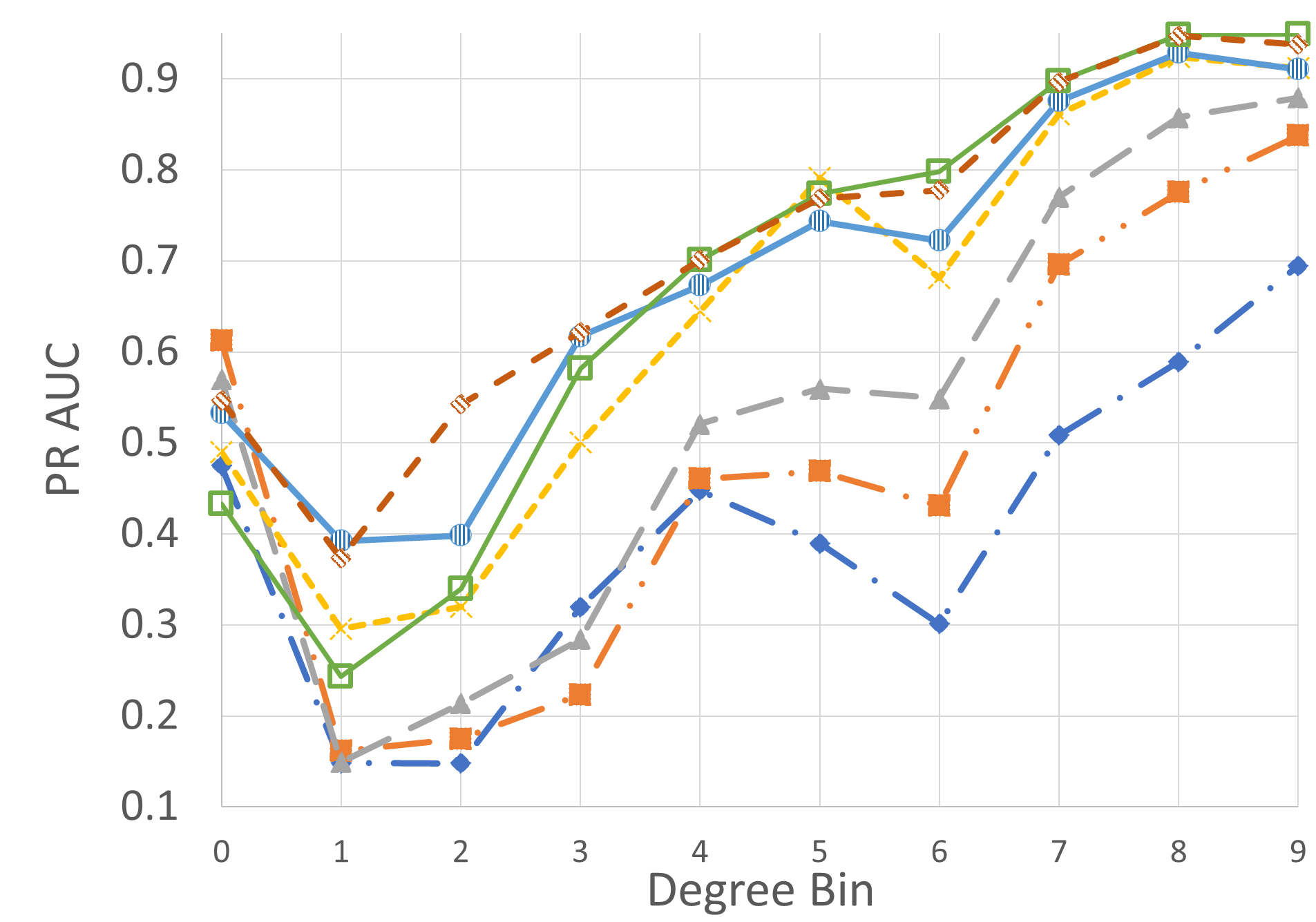}}
\caption{Performance of all methods on \drugbank under different training data ratios. Under each case, further breakdown of performance under different node degree splits (Figure~\ref{fig:degree_bin_counts}) are shown.}
\label{fig:degree_bin}
\vspace*{-6mm}
\end{figure*}

\begin{figure}[h!]
    \centering

    \subfloat[\llgnn]{\includegraphics[width=0.25\textwidth]{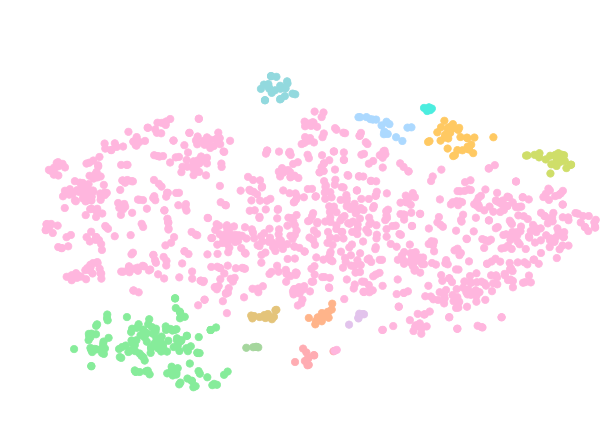}}
    \subfloat[\llgnn]{\includegraphics[width=0.25\textwidth]{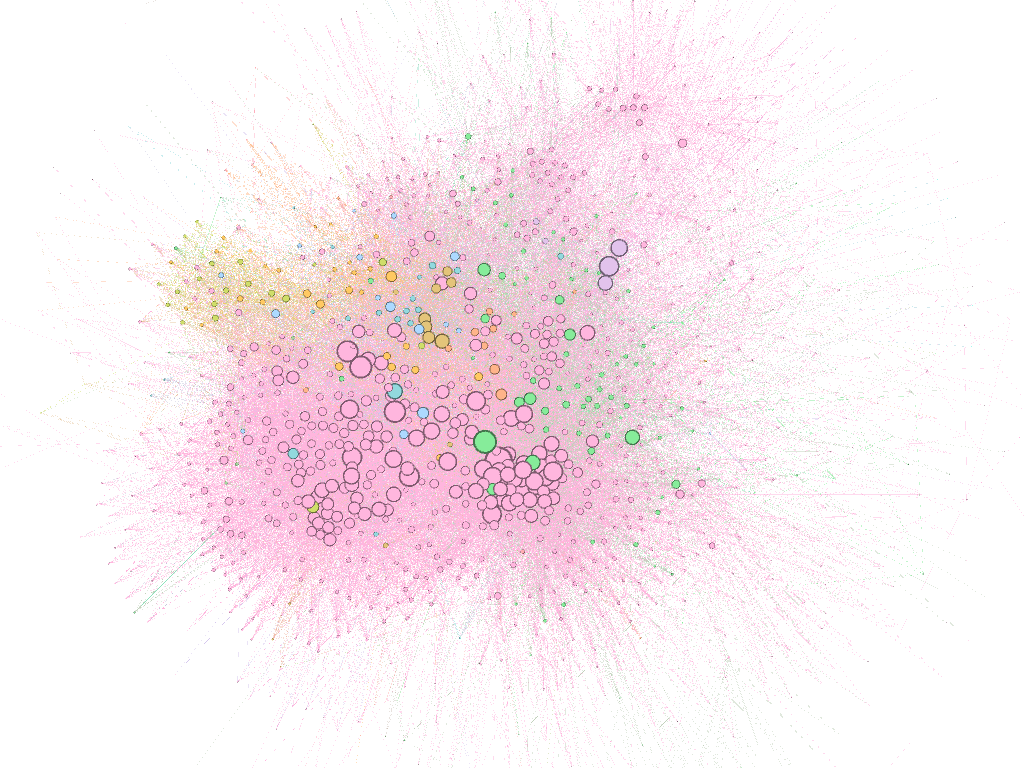}}
    \vspace*{-2mm}
    

    
    \subfloat[\decagonoh]{\includegraphics[width=0.25\textwidth]{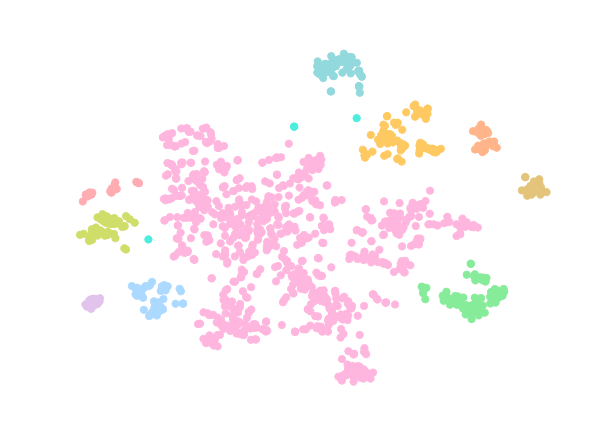}}
    \subfloat[\decagonoh]{\includegraphics[width=0.25\textwidth]{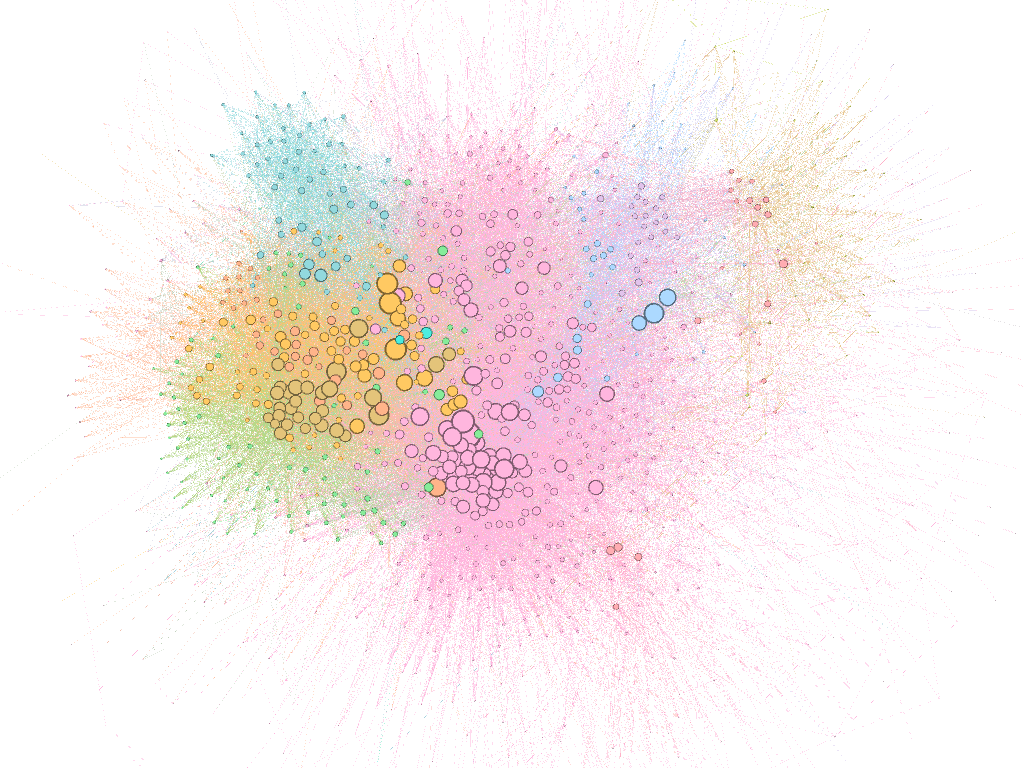}}
    \vspace*{-2mm}
    
    \subfloat[\mlgnn]{\includegraphics[width=0.25\textwidth]{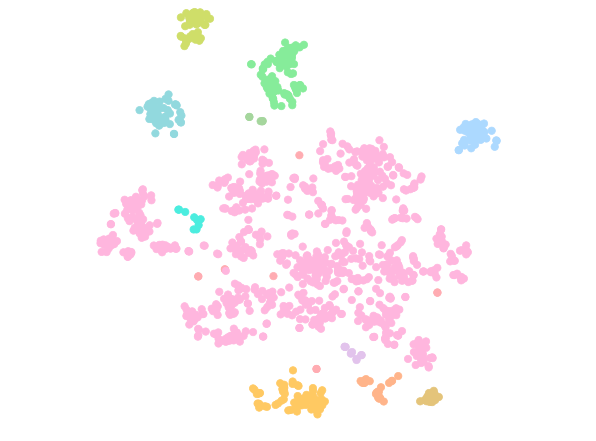}}
    \subfloat[\mlgnn]{\includegraphics[width=0.25\textwidth]{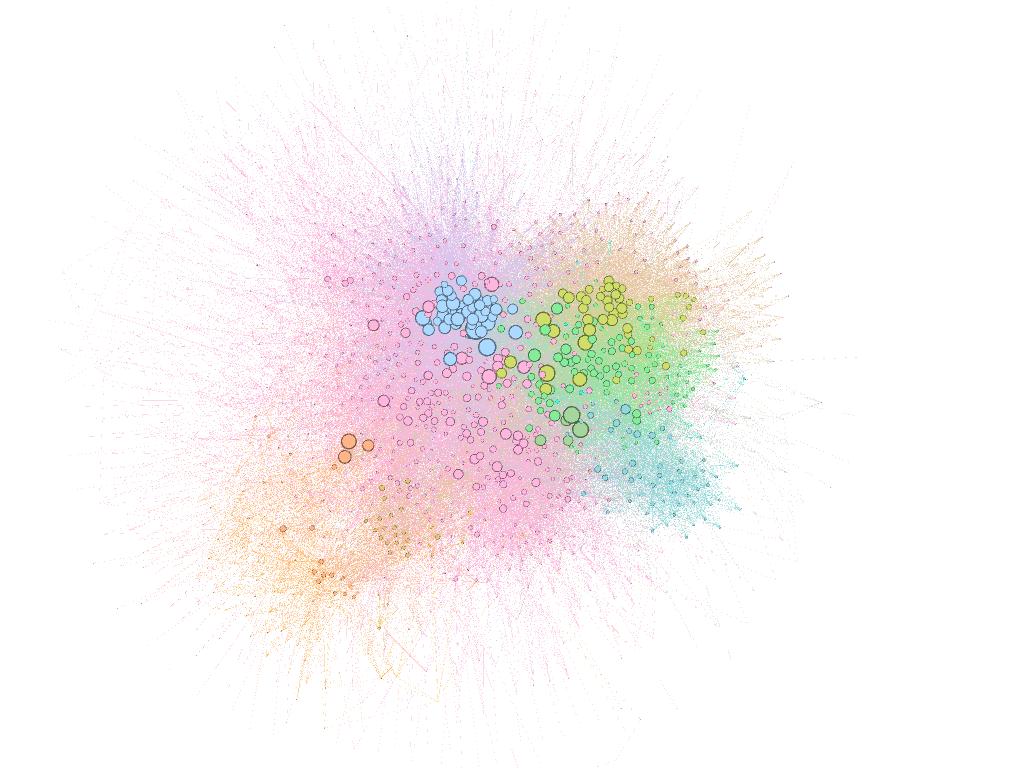}}
    \vspace*{-2mm}
    \caption{Left: Drug embeddings generated by different models with color indicating clustering. Right: The \drugbank interaction graph with node colors corresponding to the clustering and cirle size indicating node degree.}
    \label{fig:gephi}
\vspace*{-6mm}
\end{figure}

\subsection{Results}

From Table \ref{table-acc} we observe the effectiveness of different levels of network information. For models that use only the lower level representation graphs, inter-graph communication provides better performance (\llgnn vs \textsc{MHCADDI}) and utilizing GNNs allow better learning of the graph representations suit the link prediction task (\llgnn vs \fppred). However, \textsc{Decagon} baselines, which only use the higher level interaction network,  significantly outperform the lower level models. Finally, utilizing both levels of information with \mlgnn is better than \decagonfp and \decagonr and is  competitive with \textsc{Decagon-OH}. Nevertheless, it must be noted that despite the strong performance by one hot initialization, it is not scalable to larger interaction networks. Specifically, because $D^h$ grows linearly with the number of the nodes, the weight matrix becomes larger. In addition, as described in section~\ref{analysis} one hot initialization is unable to generalize representations for unseen nodes or nodes which have few neighbors.

\subsection{When \mlgnn Performs the Best}
\label{analysis}

Since the overall performance shows that \mlgnn is either the best or the second best, we conduct additional evaluation to investigate under what circumstances \mlgnn performs better than all the baseline methods. Specifically, we conjecture that the high performance achieved by \decagonoh is due to its ability to learn drug representations in a transductive way. From Equation~\ref{eq:init_transformation}, if the initial feature is one-hot, after multiplying with the learnable weight matrix $\bm{W}$, each drug is essentially represented as a randomly initialized but learnable embedding. On one hand, this inherently prohibits the using of \decagonoh to inductive setting where a drug in the testing set can be unseen during training whose embedding is then unknown. On the other hand, however, this also allows \decagonoh to take advantage of the transductive nature of the drug re-purposing task and often achieve the best performance.

What if the link structure is not available or very sparse? To \decagonoh, sparse neighborhood would limit its learning ability to adjust each drug representation, and lower performance would be expected. In contrast, our \mlgnn may still leverage the inherent molecular structure of each drug compound. Based on this hypothesis, we split the drugs into ``bins'' based on node degrees. Figure~\ref{fig:degree_bin_counts} shows the distribution of number of nodes in different bins, where bins with lower indices correspond to sparse regions of the DDI network. We then evaluate the performance of each model under each bin. As shown in Figure~\ref{fig:degree_bin}, \mlgnn indeed outperforms \decagonoh when node degree is low. In practice, drugs may come with different amount of known interacting drugs, and
our framework is especially good for drugs with little to no information available.

\subsection{Correlation between Drug Embeddings and Drug-Drug Interactions}
\label{subsec-gephi}

To gain a better understanding of what kind of representations are useful for the link prediction task, we perform the following visualization procedure. For \llgnn, \decagonoh, and \mlgnn, we take the final drug embeddings before the final prediction layer, and project them into a 2D plane using t-SNE~\cite{maaten2008visualizing}. We then use \textsc{DBSCAN}~\cite{ester1996density} to cluster the embeddings and color the drug nodes in the interaction network according to the clustering assignment, resulting in Figure~\ref{fig:gephi}. It is clear that \decagonoh and \mlgnn learn useful drug embeddings which highly correlate to the link structure in the DDI graph, while the embeddings generated by \llgnn which only relies on the molecular structure do \textbf{not} lead to good performance for link prediction in the interaction graph. Based on Section~\ref{analysis} and \ref{subsec-gephi}, we conclude that the message passing across the \textbf{higher} level graph contributes the most to the performance of GNN models on the DDI prediction task.

\section{Conclusion and Future Work}
\label{sec-conc}

We propose a multi-level GNN framework for biological entity link prediction by constructing a bi-level graph with higher level representing the interaction graph between biological entities while the lower levels representing the individual biological entity such as drug, protein, etc. In the future, we plan to extend our method to the protein-protein interaction task where each protein is represented as a hierarchical graph with amino acids, secondary structures, etc. Additionally, to improve the accuracy of our \mlgnn, we will explore the integration of graph matching and the introduction of additional input features into our framework.

\clearpage

\bibliography{bibliography}
\bibliographystyle{icml2020}

\appendix

\section{Dataset Description}
\label{sec-dataset}

We run experiments on 2 real word datasets on drug-drug interaction. Moreover, for each drug, we obtain its SMILE string from which we can derive the molecular structure. We one hot encode the lower level representation graph's node features. These include the atomic number, the number of attached hydrogen atoms, the hybridization, whether it is an acceptor or donor and whether it is aromatic.


\subsection{\drugbank}
\drugbank is a dataset of drug-drug interactions from the \drugbank database containing detailed drug data and its interaction information\footnote{\url{https://www.drugbank.ca/}}. We take the pairs mined by \cite{biosnapnets} and keep the pairs given as \drugbank Ids from which we can find the drugs' molecular structure information. After preprocessing, there are 1309 drug representation graphs and 41072 drug interactions. For training and evaluation we perform negative sampling on the positive drug pairs.


\subsection{\drugcombo}
\drugcombo is a database containing drug-drug combinations obtained from various sources including external databases, manual curations, and experimental results\footnote{\url{http://drugcombdb.denglab.org/main}}. We take the pairs of drug combinations that have been classified as exhibiting synergistic or antagonistic effects. We cross-reference PubChem\footnote{\url{https://pubchem.ncbi.nlm.nih.gov/}}, a database of chemical molecules for the molecular structures. Similar to before, we keep the pairs in which molecular structure information is found. In total, there are 3242 drug representation graphs, 34335 drug pairs classified as having synergistic effects, and 15057 pairs classified as having antongonistic effects. We also treat the different classes as edge features in the interaction graph and use it in our higher level propagation. Similar to \drugbank we perform negative sampling for each known effect.

\section{Parameter Settings}
\label{sec-detail}

For each lower level and higher level \gat we have hidden dimensions of 64. The input dimensions for \decagonfp and \decagonr are 64 dimensions while the input dimension for \decagonoh is the number of drugs in the DDI. We use 3 higher level \gat layers for \decagon models. For \llgnn we use 5 lower level \gat layers while for \coatt we use 3 blocks of \gat followed by \gmn co-attention. Experiments are ran on an Intel i7-6800K CPU and Nvidia Titan GPU. We split the dataset intro training validation and testing sets based on the known pairs. For training, for all models except \coatt, we have a batch size of 64 graphs and select all adjacent pairs in the training set. For \coatt, we have a batch size of 128 pairs as each graph is dependent on its paired graph only for its own representation. Additionally, we use the Adam optimizer~\cite{kingma2014adam} with the learning rate set to 0.001 and the the number of iterations set to 10000. We use the mean aggregator as our $\mathrm{READOUT}$ function. We evaluate on the validation set after every 100 iterations and employ early stopping if performance does not improve over a window size of 15. Finally, all experiments were implemented with the PyTorch and PyTorch Geometric libraries~\cite{Fey/Lenssen/2019}.

\section{Baseline Details and Related Work}
\label{sec-related}

\subsection{\coatt vs GMN}
\label{subsec-coatt-vs-gmn}
In this work, we mention the use of inter-graph attention in \coatt with GMNs~\cite{li2019graph}. For two graphs $\mathcal{G}^l_1$, $\mathcal{G}^l_2$ and node $\bm{x}_{i} \in \mathcal{G}^l_1$ the model as described in ~\cite{deac2019drugdrug} computes the inter-graph representation as 

\begin{equation}
\bm{x}_{i,inter}^{(k+1)} = \sum_{j \in\mathcal{G}^l_2}\alpha^{(k+1)}_{i,j}\bm{W}_1^{(k+1)}\bm{x_j}^{(k)}
\end{equation}

\begin{equation}
\alpha^{(k+1)}_{i,j} = \dfrac{\mathrm{exp}(\langle \bm{W}_2^{(k+1)}\bm{x}_i^{(k)},   \bm{W}_2^{(k+1)}\bm{x}_j^{(k)} \rangle) }{\sum_{j^{'} \in\mathcal{G}^l_2}\mathrm{exp}( \langle \bm{W}_2^{(k+1)}\bm{x}_i^{(k)},  \bm{W}_2^{(k+1)}\bm{x}_{j^{'}}^{(k)} \rangle)}.
\end{equation}

Here, $\langle \cdot, \cdot \rangle$ is the dot product. In our model, we utilize a similar attention mechanism adopted from Graph Matching Networks \cite{li2019graph}. Specifically we compute the inter-graph message as follows

\begin{equation}
\bm{x}_{i,inter}^{(k+1)} = \sum_{j \in\mathcal{G}^l_2}\alpha_{i,j}(\bm{x}_i^{(k)} - \bm{x}_j^{(k)})
\end{equation}

\begin{equation} \label{eq:gmn_att} {a}_{i,j}=  \frac {\mathrm{exp}(\mathrm{cos}(\bm{x}_{i}^{(k)}, \bm{x}_{j}^{(k)}))}{\sum_{j \in\mathcal{G}^l_2}\mathrm{exp}(\mathrm{cos}(\bm{x}_{i}^{(k)}, \bm{x}_{j}^{(k)}))}.
\end{equation}

In both cases, \coatt and GMN update the next step node representation, $\bm{x}_{i}^{(k+1)}$, using a combination of the representation from intra-graph message passing and inter-graph message passing.

\subsection{\textsc{Decagon} vs \mlgnn}

In the main text, we compare with three versions of \textsc{Decagon}~\cite{Zitnik2018}, but here we would like to point out that the original work of \textsc{Decagon} uses additional input features for the drugs as well as a heterogeneous interaction network including both drugs and proteins, whereas our three versions use the same amount of input features as \mlgnn for fair comparison. In addition, drug-drug similarity has also been explored for DDI prediction in literature~\cite{gae}. In future, we plan to incorporate these additional information to further improve the prediction accuracy.



\end{document}